\documentclass[12pt]{article}
\usepackage{graphicx}

\newcommand{\wt}{\omega_{(\gamma)} \tau }
\newcommand{\wtz}{\omega_{(1)} \tau }
\newcommand{\wto}{\omega_{(2)} \tau }
\newcommand{\half}{\frac{1}{2}}
\begin{document}

\begin{flushright}
 CECS-PHY-00/02 \\
gr-qc/0007038
\end{flushright}
\vspace{0.4cm}

\begin{center}

 {\Large {\bf   Cosmological scaling solutions of minimally coupled
 scalar fields in three dimensions}} \vskip
0.8truecm {\large Norman Cruz}
 \vskip 0.2cm {{\it Departamento de F\'{\i}sica, Facultad de
 Ciencia,\\
 Universidad de Santiago de Chile, Casilla 307, Santiago, Chile}  \\
 \small E-mail: {\tt ncruz@lauca.usach.cl}} \vspace{0.2in}

{\large Cristi\'an Mart\'{\i}nez}
\vskip 0.2cm {{\it Centro de Estudios
Cient\'{\i}ficos (CECS),\\
 Casilla 1469, Valdivia, Chile}    \\
 \small E-mail: {\tt martinez@cecs.cl}} \vskip 0.5truecm
\today
\end{center}

\vskip 0.5truecm

\begin{abstract} We examine Friedmann--Robertson--Walker models in three
spacetime dimensions. The matter content of the models is composed of a
perfect fluid, with a $\gamma$-law equation of state, and a homogeneous
scalar field minimally coupled to gravity with a self-interacting potential
whose energy density red-shifts as $a^{-2 \nu}$ , where $a$ denotes
 the scale factor. Cosmological solutions are
presented for different range of values of $\gamma$ and $\nu$. The potential
required to agree with the above red-shift for the scalar field energy
density is also calculated.
\end{abstract}

During the last years a great number of works have been dedicated to study
three-dimensional gravity, particularly after it was shown to be a soluble
system \cite{Witten} and to contain black hole solutions
\cite{BTZ}-\cite{BHTZ}.  However, previously to this results,  cosmological models 
were considered in three-dimensional gravity. In 
particular, some Friedmann--Robertson--Walker (FRW) models 
were analyzed in \cite{Giddings,Barrow,Cornish}. Recently, the cosmic holographic principle has 
been examined using these models \cite{Wang-Abdalla}-\cite{Wang-Abdalla2}. 
In this article we extend these results studying FRW models in $2+1$ dimensions filled with a
perfect fluid, obeying a $\gamma$-law equation of state, and a homogeneous
scalar field $\phi$ minimally coupled to gravity with a self-interacting
potential  $V(\phi)$. The cases where the scalar field energy density,
$\rho_{\phi}$, redshifts as $a^{-2 \nu}$,
 with $0 \leq \nu < 1$ ($a$ denotes the scale factor) are the
 only ones considered.  We find new exact solutions for some ranges of values
 of $\gamma$ and $\nu$ and the potential required to agree with the above redshift 
for the scalar field energy density is calculated.   The range of $\nu$ is chosen such that the 
scalar field contributes with a negative pressure since  we are interested in
 the counterparts of the $3+1$ dimensional ``quintessence models" 
\cite{St},\cite{CaDaSt},\cite{HuDaCaSt}. These type of models has been
investigated since cosmological observations indicate that there must be
some kind of dark energy with negative pressure in the universe. The range
considered for $\nu$ fits cosmological observations properly. Although
cosmological observations are irrelevant in three dimensions, from
theoretical point of view it is interesting to investigate the non-trivial
features of the cosmological solutions, arising from consider universes
containing two fluids, which have a positive and a negative pressure,
respectively. As we shall see below, these cases, discussed in terms of the
energy conditions, have important differences with respect to the $3+1$
dimensional ones.
Matter with negative pressure also  has  been considered in the analysis of
the holographic principle in four dimensional two-fluid universes \cite{Rama}. 

A  homogeneous and isotropic universe in three dimensions is described by
the line element
\begin{equation}
ds^2= -d\tau^2 + a^2(\tau) \left ( \frac{dr^2}{1-\mbox{$\kappa$} r^2}
     + r^2d\phi^2 \right ),
\label{RW}
\end{equation}
where $a(\tau)$ is the scale factor and $\kappa=-1, 0, 1$ for hyperbolic,
flat and circular two-dimensional spatial geometry, respectively. The
Einstein field equations of this model are
\begin{eqnarray}
\frac{\dot{a}^2+\mbox{$\kappa$}}{a^2} & = & K(\rho_m +
\rho_{\phi}),  \label{fe1} \\ \frac{\ddot{a}}{a} & = & - K(p_m +
p_{\phi})  ,\label{fe2}\\
 \ddot{\phi}\, +\,2\,\frac{\dot{a}}{a}\,\dot{\phi}\,&=&-
\,\frac{\partial{V(\phi)}}{\partial{\phi}},\label{fe3}
\end{eqnarray}
where $K$ is the gravitational constant in $2+1$ dimensions and the dot denotes a derivative with respect to $\tau$.
The energy density associated to the homogeneous scalar field $\phi(\tau)$ is
given by 
\begin{equation}
\rho_{\phi}\,=\,\frac{1}{2}\dot{\phi}^2\,+\,V(\phi)\,
\label{rhophi}
\end{equation}
and the pressure by 
\begin{equation}
p_{\phi}\,=\,\frac{1}{2}\dot{\phi}^2\,-\,V(\phi)\,.
\label{pphi}
\end{equation}
$\rho_m$ and $p_m$ denote
the density and the pressure of the fluid, respectively. 

From (\ref{fe1}) and  (\ref{fe2}) it is possible to write
\begin{equation}
\dot{\rho}_{m}+\dot{\rho}_{\phi} + 2\frac{\dot{a}}{a}(\rho_{m}+p_m +\rho_{\phi} + p_{\phi}) = 0,
\label{conservatotal}
\end{equation}
which corresponds to  {\em total} energy--momentum tensor conservation, and handling
(\ref{rhophi}) and (\ref{pphi}) it is shown that equation (\ref{fe3}) is equivalent to the conservation of the scalar field energy--momentum tensor
\begin{equation}
\dot{\rho}_{\phi} + 2\frac{\dot{a}}{a}(\rho_{\phi} + p_{\phi}) = 0.
\label{conservaphi}
\end{equation}
This occurs because the scalar field  interacts with no fields other than gravity.  Thus,  
 equations (\ref{conservatotal}) and (\ref{conservaphi}) imply 
\begin{equation}
\dot{\rho}_{m} + 2\frac{\dot{a}}{a}(\rho_{m} + p_{m}) = 0.
\label{conservafluid}
\end{equation}
Therefore, each energy-momentum tensor is conserved independently 
and the dynamics the our model is governed by the  Friedmann equation (\ref{fe1})
and the conservation laws (\ref{conservaphi}) and (\ref{conservafluid}).

We consider that the pressure $p_m$ and the density $\rho_m$ of the fluid are related by
the $\gamma$-law
\begin{equation}
p_m= (\gamma-1) \rho_m,   \qquad \mbox{with} \qquad 1 \leq \gamma
\leq 2 \,.
 \label{se}
\end{equation}
and  that the field obeys the same type of equation of state
\begin{equation}
p_{\phi}= (\nu-1) \rho_{\phi},   \qquad  0 \leq \nu < 1 .
\label{sephi}
 \end{equation}
Note that (\ref{sephi}) does not overdetermine the system of equations since the 
potential $V(\phi)$ is not specified. 
This potential,  consistent with the above equation of state, or equivalently 
with the redshift (\ref{denfield}),   will be calculated below.  This procedure was first used in
\cite{RatraPeebles}, and recently has been considered  in relation to  
quintessence scenario (see, for instance, \cite{Liddle} and \cite{delaMacorra}). 

Using the the energy--momentum tensor conservation law and the equation of state 
for each matter component, we obtain
\begin{equation}
\rho_m= \rho_{m_{0}} \left( \frac{a_0}{a} \right)^{2 \gamma},
\label{density}
\end{equation}
and
\begin{equation}
\rho_{\phi}= \rho_{\phi_{0}} \left( \frac{a_0}{a} \right)^{2 \nu},
\label{denfield}
\end{equation}
where the constants $a_0$ and $\rho_{m_{0}}$, $\rho_{\phi_{0}} $  are the scale factor and the
energy densities at  $\tau=0$, respectively.

To find  solutions for our model we replace the equations of
state (\ref{se}) and (\ref{sephi}), and the energy density expressions
(\ref{density}) and (\ref{denfield}) in equation (\ref{fe1}). After straightforward
manipulations, we obtain
\begin{equation}
\label{integra} \int_{a_0}^{a} \frac{d x}{\sqrt{K (\rho_{m_{0}} a_{0}^{ 2
\gamma}x^{ -2\gamma+2}+ \rho_{\phi_{0}} a_{0}^{ 2 \nu}x^{ -2\nu+2})
-\kappa}}=\tau \,.
\end{equation}
We note from equation  (\ref{integra}) that $\rho_{m_{0}}, \rho_{\phi_{ 0} }$ and
$ a_0$ must be satisfy
\begin{equation}
K \rho_{m_{0}}a_{0}^{2}+K \rho_{\phi_{0}}a_{0}^{2}- \kappa \geq 0,
\label{bound}
\end{equation}
 in order to have a real value for $\tau$.

Integrating left-hand side of expression (\ref{integra})  we find the
behavior of the scale factor $a$ as a function of $\tau$, with the
initial condition $a(\tau = 0)=a_{0}$. The solutions are
summarized in the tables 1 and 2.  Table 1 includes cases
with a perfect fluid with $1\leq \gamma \leq 2 $ and a
cosmological constant, since the case $\nu=0$ corresponds to a
model with cosmological constant $\Lambda\equiv K
\rho_{\phi_{0}}$. Table 2 includes the cases with a perfect fluid
and a scalar field with a negative effective pressure. Only  the
$\kappa=0$ flat case is shown for all values of $\gamma$ and $\nu$
considered.

%%%%%%%%%%%%%%%%%%%%%%%%%%%%%%%%%%%%%%
%%%%%%%%%%%%%%%%%    ------ Tabla 1 ------        %%%%%%%%%%
%%%%%%%%%%%%%%%%%%%%%%%%%%%%%%%%%%%%%%%%%
\begin{table}[t]
\caption{ Scale factor in the case  $\nu=0$. }
\vskip 3mm
\begin{tabular}{|cl|} \hline
case&\hspace{4cm}$a/a_0$ \\ \hline & \\
$ \Lambda > 0 $&
\\ $\kappa=0$,  $ 1 \leq \gamma \leq 2$&
$  ( \cosh \wt+A_{(0)}\sinh \wt)^{1 \over{\gamma}} $ \\ 
$\kappa \neq 0$,  $\gamma=1$  &$ \cosh \wtz + A_{(\mbox{$\kappa$})}\sinh \wtz  $\\ 
$\kappa \neq 0$,  $\gamma=2$&
$( (1+ B_{\mbox{$\kappa$}})\cosh \wto + A_{(\mbox{$\kappa$})}\sinh
\wto - B_{\mbox{$\kappa$}})^{1 \over 2}$ \\  &  \\
$ \Lambda  < 0 $ &
\\ $\kappa=0$, $ 1 \leq \gamma \leq 2$ &$ ( \cos \wt
 + A_{(0)}\sin \wt)^{1 \over{\gamma}}$\\ 
$\kappa \neq 0$, $\gamma=1$  &$
\cos \wtz + A_{(\mbox{$\kappa$})}\sin \wtz  $\\
$\kappa \neq 0$, $\gamma=2$ & $( (1+ B_{\mbox{$\kappa$}})\cos \wto 
+ A_{(\mbox{$\kappa$})}\sin \wto
- B_{\mbox{$\kappa$}})^{1 \over 2} $ \\  & \\
$ \Lambda = 0 $ & \\
$\gamma=1$ &$1 + ( K\rho_{m_{0}} - \mbox{$\kappa$} a_0^{-2})^{\half} \tau $  \\
$\kappa=0$, $ 1 \leq \gamma \leq 2$ &$
( 1 + \gamma (K\rho_{m_{0}})^{\half}\tau)^ {1 \over{\gamma}}$ \\
 $\kappa=1$,  $ 1 < \gamma < 2$  & $
(\cos (\gamma-1) \eta + ( K\rho_{m_{0}} a_0^{2} -1)^{\half}\sin
(\gamma-1)\eta)^{1 \over{\gamma-1}} $\\
 $\kappa$=-1$, 1 < \gamma < 2$    &$
(\cosh (\gamma-1)\eta +  ( K\rho_{m_{0}} a_0^{2} +1)^{\half}
\sinh(\gamma-1)\eta)^{1\over{\gamma-1}}$\\
$\kappa \neq 0$, $\gamma=2$& $
( 1-\kappa \tau^2+2 ( K\rho_{m_{0}} -\kappa
a_0^{-2})^{\half}\tau)^{1 \over 2} $ \\
 &  where $\omega_{(\gamma)}= \gamma | \Lambda |^{\half} $, $
A_{(\mbox{$\kappa$})}=( \frac{ \Lambda + K\rho_{m_{0}} - \mbox{$\kappa$}
a_0^{-2} }{ |\Lambda| } )^{\half}$, \\
&$B_{\mbox{$\kappa$}}= \frac{
\mbox{$\kappa$}a_0^{-2}}{2\Lambda} $\,  and $\tau=\int a(\eta) d\eta$.\\
 & \\ \hline
 \end{tabular}
 \label{table1}
 \end{table}

\vskip 0.3cm {\it Models with a cosmological constant} \vskip
0.3cm

In these cases the condition (\ref{bound}) becomes $K
\rho_{m_{0}}a_{0}^{2}+\Lambda a_{0}^{2}- \kappa \geq 0$. If 
$\Lambda = 0$ this relation implies a lower bound for
$\rho_{m_{0}}a_{0}^{2}$ given by
\begin{equation}
\rho_{m_{0}}a_{0}^{2} \geq \frac{\kappa}{K} \,. \label{mass}
\end{equation}

\begin{figure}[h]
\centering
\includegraphics[width=8cm]{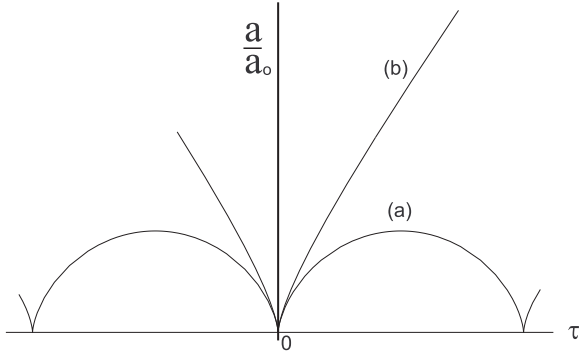}%
\caption{Scale factor $\frac{a}{a_0}$ as function of proper time $\tau$. 
(a) $\kappa=1$ closed universe and (b) $\kappa=-1$ open universe. Here 
$\Lambda=0$ and $\gamma=\frac{3}{2}$.}
\end{figure}

The solutions for $\Lambda = 0$ were analyzed in
\cite{Cornish} for radiation and dust. In the dust case
$a(\tau)\propto \tau$ independently of the spatial curvature. This occurs
since the dust matter density redshifted as $a^{-2}$ in three
dimensions and thus it can be combined with the term associated to
the curvature $\kappa$, as equation  \ref{fe1} shows. In four
dimensions, this situation occurs with texture or tangled strings,
which are particular cases of minimally coupled scalar fields with
an effective equation of state $p=- \rho/3$. They have an energy
density which redshifted as $a^{-2}$, and thus mimics a negative
curvature term. When $\kappa =1$ and $\rho_{m_{0}}a_{0}^{2}=
\frac{\kappa}{K}$ we have a special situation, which has no counterpart
in four dimensions, where the universe remains static.

In the radiation case, $\gamma =3/2$,  and
$a(\tau)\propto {\tau}^{2/3}$ if $\kappa=0$. In four dimensions, $a(\tau)\propto
{\tau}^{1/2}$ for the flat case.

The figure 1(a)  shows the scale factor for a closed universe $a$ reaches a 
maximum value and   contracts to zero in a finite proper time. 
The figure 1(b) shows the case $\kappa=-1$ open universe. This universe expands
forever after  $a$ vanishes. 

\begin{figure}[h]
\centering
\includegraphics[width=8cm]{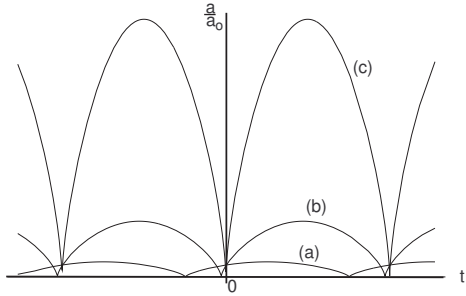}%
\caption{Scale factor $\frac{a}{a_0}$ as function of $t=\frac{3}{2}  \sqrt{\Lambda}  \,\tau$.
 Universes  with negative cosmological constant and $\kappa=0$. 
(a) $A_{(0)}=0.1 $ (b) $A_{(0)}=10$ (c) $A_{(0)}=100$. Here $\gamma=\frac{3}{2}$.}
\end{figure}

If  $\Lambda < 0$, the solutions with $\kappa=0$, $1 \leq \gamma \leq 2$
and  $\kappa \neq 0$, $\gamma=1$ always collapses to $a=0$ in a finite
proper time as is shown in figure 2. For the dust case, the proper time is given by
\begin{equation}
\tau = \frac{1}{| \Lambda |^{\half}} \arctan \left ( \left (
\frac{K\rho_{m_{0}} a_0^{2}-\kappa - |\Lambda|a_0^{2}}{|\Lambda|a_0^{2}}
\right )^{-1/2} \right ),
 \label{tiempo}
\end{equation}
which was found in \cite{Mann}, in order to prove that static
black hole solution in 2+1 dimensions arises naturally from
gravitational collapse of pressureless dust with a negative
cosmological constant. Since $\Lambda < 0$ implies an attractive
cosmological force these universes begin in a singularity and end
in a big crunch, i. e., they behave like the standard closed
model, even with a flat or hyperbolic curvature. 

\begin{figure}[h]
\centering
\includegraphics[width=8cm]{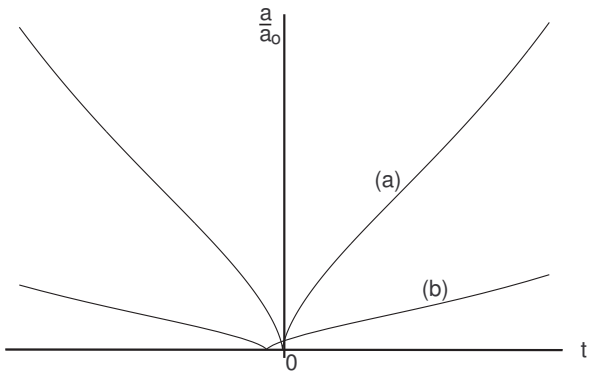}
\caption{Scale factor $\frac{a}{a_0}$ as function of $t=\frac{3}{2}  \sqrt{\Lambda}  \,\tau$. Universes 
 with positive cosmological constant and $\kappa=0$. 
(a) $A_{(0)}=100$ (b) $A_{(0)}=10$. Here $\gamma=\frac{3}{2}$.}
\end{figure}

If  $\Lambda > 0$, the solutions for universes filled with dust,
are always inflationary-type solutions and the scalar field vanishes at
\begin{equation} \label{tauv}
\tau=\frac{1}{2 \sqrt{\Lambda}} \ln \frac{ A_{(\kappa)}-1}{A_{(\kappa)}+1}
\end{equation}
 if $A_{(\kappa)}>1$, that is, if $ K\rho_{m_{0}}> \kappa a_0^{-2}$. 
In the flat case and $1 <\gamma \le 2$, there are no restrictions and $a$ 
always vanishes at a proper time given by a expression similar to (\ref{tauv}).

\vskip 0.8cm {\it Models with scalar field} \vskip 0.3cm

From the expression $ \dot{\phi}^2=\rho_{\phi}+p_{\phi}= \nu\rho_{\phi}$, and
the Friedmann equation (\ref{fe1}), we obtain 
\begin{equation}
\frac{d\phi}{da}=\frac{1}{a} \sqrt{\frac{ \nu
\rho_{\phi}}{K(\rho_{\phi}+\rho_{m})}}
\label{phivsa}
\end{equation}
for the flat case $\kappa=0$. After replacing equations  (\ref{density}) 
and (\ref{denfield}) into above equation, we do the integration using
\begin{equation}
\int\frac{dx}{x
\sqrt{c^2+x^{-n}}}=\frac{2}{cn}\mbox{arcsinh} \,{c}\, { x^{\frac{n}{2}}}.
\end{equation}
Thus, we  obtain an explicit expression of
$a$ in terms of $\phi$. Now, considering the relation $V(\phi)=(1-\nu/2)
\rho_{\phi}$ and equation (\ref{denfield}), we write down the potential
\begin{eqnarray}
V(\phi)&=&(1-\nu/2) \rho_{\phi_{0} }[(\cosh (\gamma-\nu)\sqrt{
\frac{K}{\nu}}(\phi-\phi_{0}) \nonumber\\ & + &\displaystyle (1+ \frac{
\rho_{m_{0}}}{\rho_{\phi_{0} } } )^{\half}\sinh(\gamma-\nu)\sqrt{
\frac{K}{\nu}}(\phi-\phi_{0})]^{{-2 \nu} \over{\gamma-\nu}} . \label{pot}
\end{eqnarray}

\begin{table}
\caption{Scale factor in the cases $\nu \neq 0$}
\vskip 3mm
\begin{tabular}{|cl|} \hline
case&\hspace{4cm} $a/a_0$ \\ \hline & \\
$\begin{array}{c}  \kappa = 0 \\  0<\nu<1\\ 1 < \gamma < 2 \end{array}$  & $\begin{array}{l}
 \qquad \begin{array}{r}
[\cosh \beta ( \eta-\eta_{0})  + (1+ \frac{ \rho_{\phi_{0}}}{\rho_{m_{ 0}} }
)^{\half}\sinh\beta ( \eta-\eta_{0})]^{1 \over{\gamma-\nu}}
 \end{array} \end{array}$ \\  
& \qquad  where $\beta=(\gamma-\nu)\sqrt{ K\rho_{\phi_{0}}}a_{0}^{\nu}$  
 and $\tau=\int a^{\nu} d\eta$,\\
 $\begin{array}{c} \kappa = -1,0,1 \\ \nu=1/2 \\ \gamma=1 \end{array}
$ & $\begin{array}{l}
 \qquad  \begin{array}{rl}
 \frac{ K \rho_{\phi_{0}} \tau^2}{ 4} + \left
 [K(\rho_{m_{0}}+\rho_{\phi_{0}} )
 - \kappa a_0^{-2} \right]^{1/2} \tau + 1
  \end{array} \end{array}$ \\  & \\
\hline
 \end{tabular}
 \label{table2}
 \end{table}

In order to consider the model where only a scalar field energy density
exists, we set $\gamma=1$ and $\rho_{m_{0}}=0$. From (\ref{pot}) we read
the field potential in this case
\begin{equation}
V(\phi)=(1-\nu/2) \rho_{\phi_{0} }\exp(-2 \sqrt{K \nu}(\phi-\phi_{0})).
 \label{pot00}
\end{equation}
With this potential we find a special solution for which 
$$ a(\tau)=a_0
[1+\nu \sqrt{K \rho_{\phi_{0} }} \tau]^{1/\nu} \quad \mbox{and} \quad
\phi(\tau)=\frac{1}{\sqrt{K\nu}} \ln|1+\nu\sqrt{K\rho_{\phi_{0}}}\tau |+\phi_0 \,.$$

In \cite{Barrow} this potential, with $\nu=1/2$, was used and a different solution
was found (but with the 
same asymptotic behavior).  This solution does not belong 
to the class of solutions  that satisfies $p_{\phi}/\rho_{\phi}=\mbox{constant}$.

Is interesting to observe that in the cases with
$\nu=1/2$ and dust as matter component, we obtain an inflationary
power-law solution.
\begin{figure}
\centering
\includegraphics[width=8cm]{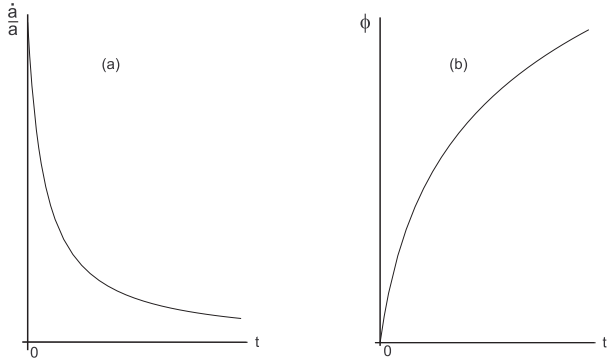}%
\caption{(a) The expansion rate of the cosmological model which only contains
 a scalar field only with $\nu=1/2$.  (b) The time dependence of the scalar
 field. Here $t=\sqrt{K \rho_{\phi_0}} \tau$ and we set $\phi_0=0$.}
\end{figure}

The cosmological solutions  found in this paper can be interpreted from the
point of view of the specific energy conditions that guarantee geodesic
convergence under gravity in $2+1$ dimensions (see Barrow {\it et al}
\cite{Barrow}). The {\it weak energy condition} (WEC) implies, for a perfect
fluid, regardless of the number of spatial dimensions, the condition
$\rho\,>\,0$, that is, the matter density must be positive. The {\it strong
energy condition} (SEC) in a $(D+1)$-dimensional general relativistic
spacetime leads to the following restriction on $\rho$ and $p$
\begin{equation}
(D-2) \rho + D p > 0 \,.
\label{SEC}
\end{equation}
In $2+1$ dimensions equation (\ref{SEC}) implies that the pressure must be
positive, $p\,>\,0$. In the $(3+1)$-dimensional case SEC implies $\rho +
3p\,>\,0$. Therefore, in $2+1$ dimensions WEC and SEC place independent
positivity conditions on the density and pressure. Equation (\ref{fe2})
allows us to conclude that if SEC is satisfied (in $2+1$ dimensions) then
\cite{Visser}
\begin{equation}
\ddot{a}\,<\,0,
\label{acelera}
\end{equation}
which is analogous to the $(3+1)$-dimensional case. The corresponding
cosmological solutions must be non-inflationary, independent of whether the
universe is open, flat, or closed. SEC can be violated by a positive
cosmological constant or by a a scalar field with a negative mean pressure 
\cite{Visser1}. In the models of inflation in $3+1$ dimensions, which
consider an effective cosmological constant, all the solutions have
$\ddot{a}\,>\,0$. A late-time accelerated expansion of the scale factor
occurs in FRW models filled with dust and some kind of energy with a
negative average pressure (a positive cosmological constant or a self-
interacting scalar field). In these scenarios, during the most part of
the lifetime of the universe  the expansion is decelerating. At very late times
the acceleration becomes zero and then begins to increase. 

In the $(2+1)$-dimensional case, since the acceleration depends only on the
pressure, an inflationary solution for the scale factor is obtained if the
total pressure is negative. For a dust filled universe with a 
negative cosmological
constant or a scalar field energy density the scale factor is always
accelerating. Solutions with a late-time accelerated expansion are found for
perfect fluids with nonzero pressure, i. e., with $1\,<\,\gamma \leq 2$.

\section*{Acknowledgements}
This work was partially supported by the grants Nos. 3970004 from FONDECYT
(Chile) and 04-9331CM from DICYT, Universidad de Santiago de Chile. The
institutional support to the Centro de Estudios Cient\'{\i }ficos de
Santiago of Fuerza A\'{e}rea de Chile, I. Municipalidad de Las Condes, I.
Municipalidad de Santiago and a group of Chilean companies (AFP Provida,
Codelco, Copec, Empresas CMPC, Gener S.A., Minera Collahuasi, Minera
Escondida, Novagas, Business Design Associates, Xerox Chile) is also
recognized. CECS is a Millennium Science Institute.

\end{document}